\documentclass[journal=jacsat,manuscript=article]{achemso}

\usepackage[version=3]{mhchem} 

\usepackage{braket}

\usepackage{hyperref}

\usepackage{setspace}

\usepackage{adjustbox}

\usepackage{color}

\newcommand{\duo}{{\sc Duo}}
\newcommand{\Duo}{{\sc Duo}}

\newcommand{\ai}{\textit{ab initio}}

\author{Ryan P. Brady}
\email{ryan.brady.17@ucl.ac.uk}
\affiliation{Department of Physics and Astronomy, University College London, Gower Street, WC1E 6BT London, United Kingdom}
\author{Sergei N. Yurchenko}%
 \email{s.yurchenko@ucl.ac.uk}
\affiliation{Department of Physics and Astronomy, University College London, Gower Street, WC1E 6BT London, United Kingdom}

\date{\today}

\title{Spin-Orbit Induced Non-Adiabatic Dynamics: An Exact $\Omega$-Representation}

\abbreviations{IR,NMR,UV}

\begin{document}








\begin{abstract}

Transforming  rovibronic Hamiltonians of molecular systems from the $\Lambda S$ (Hund’s case a) basis to the adiabatic $\Omega$ representation is widely used to ``remove'' spin-orbit coupling (SOC) and enable single-state treatments of spectra and dynamics. We show that this simplification is only apparent: the SOC elimination necessarily generates sizeable non-adiabatic couplings (NACs) from the nuclear kinetic energy operator. Neglecting these spin-orbit-induced NACs causes severe errors in rovibronic energies and transition properties. Using an analytically tractable two electronic state model and high-accuracy variational benchmarks, we derive the exact conditions for numerical equivalence between $\Omega$ and $\Lambda S$ formulations and quantify how missing NAC terms and bond-length-dependent spin factors degrade predictions. We implement a complete $\Omega$-representation workflow in Duo for diatomics, fully transforming all Hamiltonian terms and enabling side-by-side $\Omega$ vs $\Lambda S$ calculations. For common single-state pipelines (e.g., LEVEL), we provide diagnostics that flag unsafe regimes and practical remedies to restore accuracy. The results deliver actionable guidance for spectroscopy, photophysics, and kinetics: $\Omega$-based single-state approximations are reliable only when interacting states are well separated in the Franck-Condon region; otherwise, explicit non-adiabatic terms are required—even for ``forbidden'' transitions.

\end{abstract}

\maketitle


\section{Introduction}
\label{sec:Intro}

Relativistic SOC is a fundamental interaction in molecular systems, giving rise to the fine structure of rovibronic energy levels, reaction pathways between otherwise non-interacting states \citep{18PeGiDa.omega,14MaMaGo.omega} (e.g. interstate crossings between singlets and triplets), and making dark states bright through the intensity stealing mechanism \citep{22BrYuKi.SO,22BeJoLi.SO}. Of importance in ultracold molecular physics is exactly this SOC which enables optical access to metastable states that are otherwise spin-forbidden from the ground state. The resulting transitions are extremely narrow ($\sim10^0-10^2$ kHz natural linewidths), providing long-lived states for precision spectroscopy and coherent optical control of ultracold molecules \citep{14KoAiOa.omega,26MuTo.diabat,10TaKlKr.KCs,11ToPaJe.SrYb,12ToGoMu.Rb2,16AlKrTa.KRb,98BeThSh}. SOC splits rovibronic states into distinct $\Omega$ fine-structure components, and a spectroscopic Hamiltonian that explicitly incorporates these $\Omega$-states provides a natural framework for tuning dynamics and enabling controlled state transfer in ultracold molecules.



Unitary transformations are a powerful tool to simplify complex quantum-systems, such as molecules, by transforming their Hamiltonian into a more convenient representation. An established example is the adiabatic to diabatic transformation (AtDT) \citep{00Baer.diabat, 00BaAlxx.diabat,02BaerMichael.diabat, 06Baer, 69Smith.diabat,19YaXiZh, 18VaPaTr.diabat, 20MaZrBe.diabat,81Delos.diabat, 82MeTrxx.diabat, 04JaKeMe.diabat,89Baer.diabat,24BrDrYu.diabat,22BrYuKi.SO, 25Brady.diabat} which attempts to remove all (radial) non-adiabatic couplings (NACs) arising from the decoupling of electronic and nuclear motion in the Born-Oppenheimer approximation \citep{27BoOpxx.diabat,54BoHuxx.diabat}. This transforms the system from a complex adiabatic representation, with cusps in its potential energy curves (PECs) and a non-diagonal kinetic energy matrix containing NACs \citep{82MeTrxx.diabat,04JaKeMe.diabat,18KaBeVa.diabat,21ShVaZo.diabat}, into a simpler diabatic representation featuring smoother, crossing PECs and a diagonal kinetic energy at the cost of introducing off-diagonal diabatic potential couplings (DCs) \citep{82MeTrxx.diabat, 04JaKeMe.diabat,81Delos.diabat}. The AtDT is known to be an important tool for molecular physicists as it allows the choice of representation for different molecular systems, of any size  \citep{03ByVoMa.thesis, 01Schwenke.thesis,15SuMeRa.thesis,19YaXiZh,82MeTrxx.diabat,22ShVaZo.diabat,02AbKuxx.diabat,02AbKuxx.diabat,06Baer,24BrDrYu.diabat,25Brady.diabat,25BrYuxx.diabat}. For example, charge transfer effects in molecules can be mediated by the associated NACs \citep{92BoGaxx.thesis, 95BeGaxx.thesis,10VoKoBe.thesis}, for explaining complex vibrational progressions in the spectrum of molecules (e.g. the Clements band of SO2 \citep{13XiHuZh.SO2}), and in understanding the UV intensity problem of water \citep{03ByVoMa.thesis}. This list is non-exhaustive, but underscores the importance of non-adiabatic effects in high accuracy sciences such as high resolution spectroscopy.

Similarly to the AtDT, the $\Lambda S$ (Hund's case (a)) to $\Omega$-basis transformation has been widely used \citep{21BeJoLi.SO,19PaEvAn.ai, 00BeScWe.ai,11YuBixx.SO,25RoGoDi.diatom} to simplify the treatment of spin-orbit coupling (SOC) in rovibronic calculations.  The transformation to the $\Omega$-representation, known as the state interacting method \citep{19PaEvAn.ai, 00BeScWe.ai, 11YuBixx.SO,25RoGoDi.diatom}, aims to remove the SOC by diagonalisation of the Breit-Pauli spin-orbit (SO) Hamiltonian ($\mathbf{H}_{\rm SO}$) together with the electronic Hamiltonian, yielding effective potentials for each SO-component. 

This procedure of removing SOC has the attractive motivation of electronically decoupling the system and transforming it to a single state representation, greatly simplifying the solution of the nuclear motion problem. However, like the AtDT, transforming to the $\Omega$-representation (``adiabatic'') unavoidably introduces non-adiabatic effects that are essential to include for treating general spectroscopic systems accurately. 
Despite the known importance of properly treating these non-adiabatic effects in some diatomic studies \citep{10TaKlKr.KCs,10KrKlNi.KCs}, the simplified, fully decoupled $\Omega$-representation remains popular in spectral applications, especially for calculating transition properties of forbidden bands, such as intensities or lifetimes. Some (non-exhaustive) examples include Refs. \citep{12SrSa.CH-,12XiShSu.SO+,14ShLiSu.NSe,19ChElEl.YbBr,19YaYiLi.AlI,19LiLiLi.SiO+,19AbElKo.MgCl+,21XiLiRe.PSe,21ZeElFa.BeSe,22BeJoLia.SO,23ElAlAb.LuF,24AkZrLa.BaLi+,24XiWeAn.PO,24MoNoMa.AgH}. A common assumption is that these properties are less sensitive to non-adiabatic effects than rovibronic energies, which are often empirically refined in high-resolution studies. Due to a lack of accurate experimental data for intensities or lifetimes, transition dipoles are often treated \ai\ while the single-state approximations remain untested. Single-state representations are desirable because they can be easily used with readily available single-state methods, such as LeRoy's \textsc{Level} program \citep{LEVEL}, to compute transition intensities  \citep{21BeJoLi.SO} for complex multi-state systems, including forbidden bands. Not only in spectroscopy, but the SO-induced NACs are pivotal in understanding spin-transfer processes in molecules, much like charge transfer dynamics are mediated by electronic-nuclear NACs.

In this work, we rigorously assess the $\Omega$-representation for a diatomic system since the theory can be exactly analysed and high precision science (e.g. ultracold molecular physics) is typically limited to studying small molecules. We quantify the impact on the rovibronic solution when omitting SOC-induced NAC terms, focussing on dipole forbidden transition intensities where these effects are expected to be important. Our findings reveal, contrary to its reputation as a simplified but reasonably accurate transformation, the single state approximation commonly used with the $\Omega$-representation may introduce errors in transitions intensities (and radiative lifetimes) that make it unreliable for spectroscopic applications. Furthermore, non-adiabatic couplings in the $\Omega$-representation makes it less attractive than the $\Lambda S$ treatment. We also discuss system types and use cases where the single-state approximation in the $\Omega-$representation is expected to be sensible when a full treatment of the non-adiabatic effects is intractable.


\section{Transforming to the $\Omega$ Representation of Nuclear Motion}
\label{sec:theory}
To transform from the $\Lambda S$ to the $\Omega-$representation, one needs to follow the general method:

\begin{itemize}
    \item Solve the electronic Schr\"{o}dinger equation for electronic wavefunctions, and subsequently construct potential energy and SOC surfaces for the electronic states of interest.
    \item Build, in the electronic basis, the electronic + Breit-Pauli Spin-Orbit Hamiltonian matrix: $\mathbf{H_{\Omega}=}\mathbf{V}+\mathbf{H}_{\rm SO}$ at every nuclear configuration grid point.
    \item Diagonalise $\mathbf{H_\Omega}$ at every nuclear configuration to obtain effective Spin-Orbit decoupled states and potential energy surfaces.
    \item The single state approximation finishes at the previous point. But now we need to unitarily transform the entire nuclear motion Hamiltonian by the diagonalising transformation from above at every nucelar geometry, i.e. the vibrational and rotational kinietic energy Hamiltonians with their respective derivative operators, and the remaining coupling terms such as electronic angular momentum. This will produce non-adiabatic coupling terms which are necessary to achieving exact equivalence with the original $\Lambda S$ representation. 
\end{itemize}
We now move to a diatomic molecule, where the $\Omega$ transformation is demonstrated.

\subsection{The Diatom}

Let us start with the $\Lambda S$ (Hund's case (a)) basis representation, where the spin-orbit-coupled, rovibronic Schrödinger equation for a diatomic molecule reads
\begin{align}
\label{eq:diatomic_schrodinger_equation_duo}
\left[  \frac{\hbar^2}{2\mu} \left(-\frac{d^2}{dr^2} + \frac{1}{r^2} \hat{\mathbf{R}}^2\right) + \mathbf{V} + {\mathbf{H}}_{\rm SO} \right] \vec{\chi} = E_i \vec{\chi}.
\end{align}
Here, $\mu$ is the reduced mass, $r$ is the internuclear separation, $\vec{\chi}$ is the rovibronic wavefunction vector, and $E_i$ is the corresponding energy eigenvalue. The first and second terms are the vibrational and rotational kinetic energy operators, respectively. The matrix $\mathbf{V}$ contains diagonal Born-Oppenheimer potential energy curves \citep{27BoOpxx.diabat,54BoHuxx.diabat} (PECs). The SO  electronic matrix elements above can be obtained \textit{ab initio}, e.g., using \textsc{Molpro}, and are used to construct the full Hamiltonian in Eq.~\eqref{eq:diatomic_schrodinger_equation_duo}.
This Hamiltonian can be solved variationally (we use  our nuclear motion code \textsc{Duo} \citep{Duo}).


Rather than solving Eq.~\eqref{eq:diatomic_schrodinger_equation_duo} directly, let us follow the the state-interacting approach of the $\Omega$-representation  \citep{21BeJoLi.SO,19PaEvAn.ai,00BeScWe.ai}, in which we eliminate spin-orbit coupling via the following transformation at each bond length $r$
\begin{align}
\label{eq:diagonalisation}
    \mathbf{U}^\dagger(r) \left( \mathbf{V}(r) + \mathbf{H}_{\rm SO} \right) \mathbf{U}(r) = \mathbf{V}_{\Omega}(r),
\end{align}
where $\mathbf{U}(r)$ is the $r$-dependent unitary transformation to the $\Omega$ basis, and $\mathbf{V}_\Omega(r)$ is a diagonal matrix of spin-orbit-decoupled PECs labelled by the projection of total electronic angular momentum, $\Omega$. The transformation of the $\Lambda S$ electronic basis is then 
\begin{align}
\label{eq:omega_rovibrational_basis}
 \ket{\text{state},\Lambda,S,\Sigma} \rightarrow  \ket{\text{state},\Omega}.
\end{align}

It is tempting to assume that this transformation completely decouples the $\Omega$ states, allowing the rovibronic problem to be reduced to a single-state problem. However, this assumption is (strictly speaking) incorrect (see, e.g. Ref.~\citenum{10TaKlKr.KCs}). To remain consistent, the full rovibronic Hamiltonian of Eq.~\eqref{eq:diatomic_schrodinger_equation_duo}, including the vibrational and rotational kinetic energy operators, must also be transformed. Crucially, the transformation of $\frac{d^2}{dr^2}$ in Eq.~\eqref{eq:diatomic_schrodinger_equation_duo} introduces NACs. Following standard diabatisation theory  \citep{00Baer.diabat,00BaAlxx.diabat,02BaerMichael.diabat,06Baer,69Smith.diabat,18VaPaTr.diabat,20MaZrBe.diabat,24BrDrYu.diabat,25Brady.diabat}, transformation of the vibrational kinetic energy yields
\begin{align}
\label{eq:SO_NACs}
&-\frac{\hbar^2}{2\mu}\bra{\chi_i}\mathbf{U}^\dagger \overrightarrow{\frac{d^2}{dr^2}} \mathbf{U}\ket{\chi_j} \notag \\
&= -\frac{\hbar^2}{2\mu} \bra{\chi_i} \left[ \overrightarrow{\frac{d^2}{dr^2}} + \mathbf{W}^2 - \left( \overleftarrow{\frac{d}{dr}} \mathbf{W} - \mathbf{W} \overrightarrow{\frac{d}{dr}} \right) \right] \ket{\chi_j},
\end{align}
where $\mathbf{W}(r) = \mathbf{U} \, d\mathbf{U}^\dagger/dr$ is a skew-Hermitian matrix of derivative couplings. The diagonal elements of $\mathbf{W}^2$ act as perturbative corrections to the potential energy curves, analogous to the diagonal Born-Oppenheimer correction (DBOC), while the off-diagonal terms mediate nonadiabatic transitions between states of the same $\Omega$.  

Our recent work \citep{24BrDrYu.diabat,25Brady.diabat} demonstrates that neglecting NACs in adiabatically coupled  systems can lead to substantial errors in both rovibronic energy levels and wavefunctions (and therefore intensities). Their inclusion is therefore essential for accurate modelling. 

The $\Omega$ representation may be viewed as the analogue of the adiabatic representation in molecular electronic structure, where electronic and nuclear degrees of freedom are decoupled at the cost of introducing derivative couplings (DDRs). By contrast, the $\Lambda S$ basis functions as a diabatic representation: the kinetic energy remains diagonal, while nonadiabatic effects enter through off-diagonal elements of the potential, such as spin-orbit coupling. A unique feature of spin-orbit-induced NACs is that they can couple states of different symmetry but same $\Omega$, introducing additional complexity relative to traditional Born-Oppenheimer NACs.
As we will show in our toy model systems, non-adiabatic effects associated with the $\Omega$ representation  can be also significant and should not be blindly ignored.

\section{Spectroscopic Model}
\label{sec:spectroscopic_model}

\begin{figure}[tbp!]
    \centering
    \includegraphics[width=0.7\linewidth]{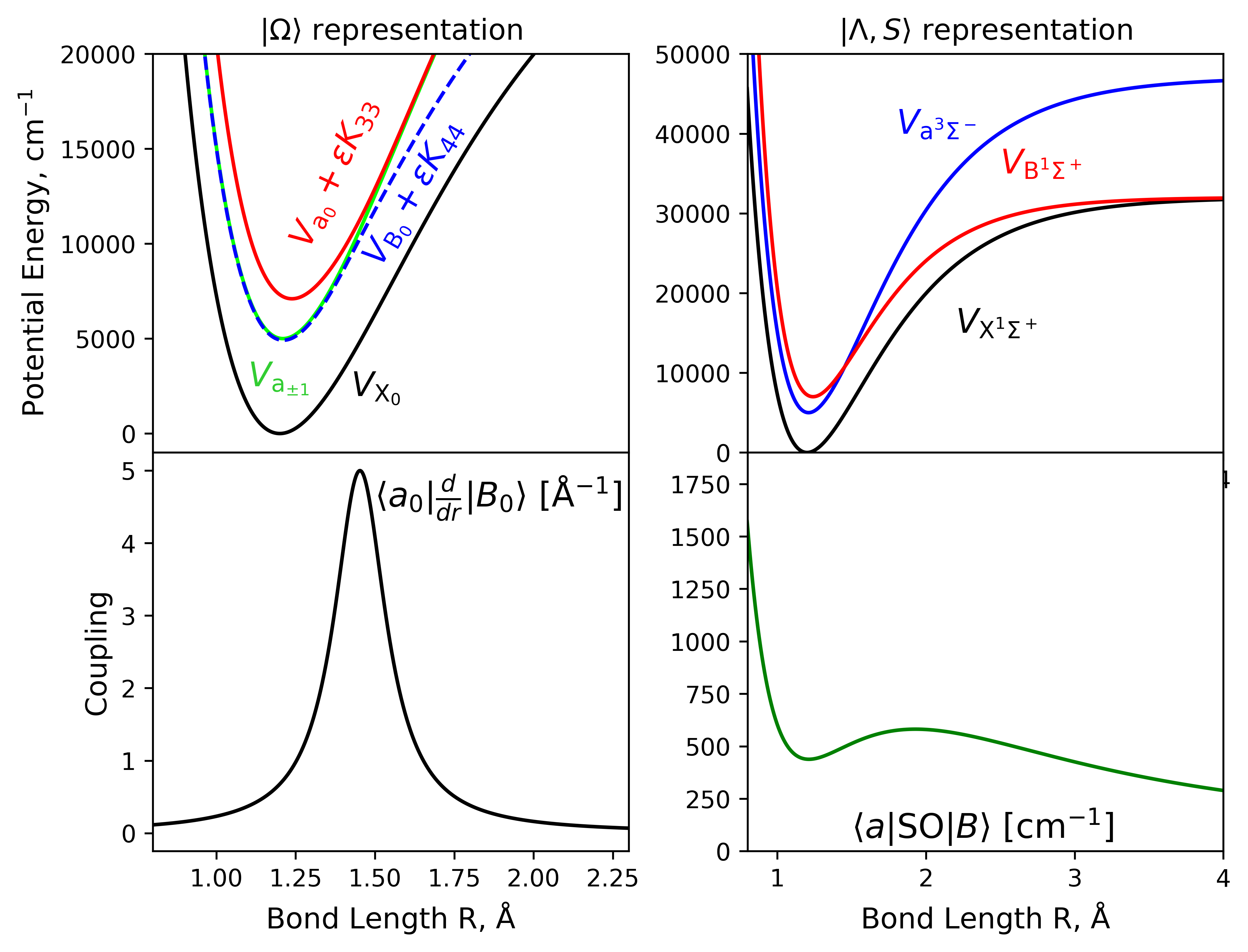}
    \caption{Illustration of the potentials (top panels) and associated couplings (bottom panels) of the two-state coupled system in the $\Omega$-representation (left panels) and $\Lambda S$ representation (right panels). The DBOC-like corrections have been added to the $\Omega$ potentials.
    }
    \label{fig:Omega_LS_model_comp}
\end{figure}

In order to investigate the influence of non-adiabatic effects on a forbidden electronic band in the $\Omega$ representation, we constructed a toy spectroscopic model as a three-state, triplet-singlet electronic system consisting of a ground  electronic state, $X\,^1\Sigma^+$ and two upper electronic states, $a\,^3\Sigma^+$ and $B\,^1\Sigma^+$ coupled by a SOC, shown in Fig.~\ref{fig:Omega_LS_model_comp}. The electronic transitions are mediated by a single transition  dipole moment curve (TDMC) between the $X\,^1\Sigma^+$ and $B\,^1\Sigma^+$ states (see Supplementary \citep{supplement} for a detailed description of the spectroscopic model). This is a simple spin-orbit system giving rise to only three spin components ($\Omega = -1, 0$ and 1). 
Two excited bound potential energy curves intersect (the $\Lambda S$ representation) near their minima which ensures that the $\Omega$-transformation significantly impacts low-lying bound states. 
This is also a typical example of a forbidden electronic band $a-X$ induced by a spin-orbit interaction between $a$ and $B$ via an intensity stealing from the dipole allowed $B-X$ band, see, e.g. Refs.~\citenum{10TaKlKr.KCs,11ToPaJe.SrYb,12ToGoMu.Rb2,16AlKrTa.KRb} as well as  the recent work by \citet{26MuTo.diabat} on several alkali-metal diatomics.  The small separation between the $X$ state and the  $a$/$B$ system is chosen to make our calculations more compact and thus easier to analyse. 


Using our nuclear motion code \duo\ \citep{Duo}, we solve the Hamiltonian of Eq.(\ref{eq:diatomic_schrodinger_equation_duo}) and perform spectral calculations in two  representations, $\Lambda S$ and $\Omega$ and hence establish a numerical equivalence of these representations for the exact $\Omega$ representation case. Using this exact solution as a reference, we  will be able to apply and analyse the impact of different approximations. 



All curves are given as analytic functions in the $\Lambda S$ representation to reduce numerical errors and are available in the supplementary material.

\subsection{$\Omega$ representation}

 
The $\Lambda S$ PECs and SOC are transformed using Eq.~(\ref{eq:diagonalisation}) to obtain spin-orbit decoupled $\Omega$-states $X\,^1\Sigma^+_0$, $B\,^1\Sigma^+_0$, $a\,^3\Sigma^-_0$, and $a\,^3\Sigma^-_{\pm 1}$, the NAC  $\mathbf{W}(r)$ and DBOC $\mathbf{W}^2$ terms. Figure \ref{fig:Omega_LS_model_comp} illustrates these curves, which form avoding crossing between states of the same $\Omega$,  $B\,^1\Sigma^+_0$, $a\,^3\Sigma^-_0$.   The DBOC-like diagonal corrections to the $\Omega$-potentials have been added to these curves and are seen to produce a minimal barrier. However, stronger NAC systems (and thus DBOCs) have been shown to produce huge spike-like barriers \citep{24BrDrYu.diabat,25BrYuxx.diabat,25Brady.diabat}, and so should be checked. It is therefore expected these SO-induced NAC terms can be important in the final rovibronic solution.



Apart from the derivative couplings $\mathbf{W}(r)$ and $\mathbf{W}^2$ produced from the transformation in Eq.~\eqref{eq:SO_NACs} using $\mathbf{U}(r)$, the rotational kinetic energy operator 
\begin{align}
\label{eq:diatomic_rotational_schrodinger_equation_duo}
\mathbf{H}_{\rm rot} = \frac{\hbar^2}{2\mu r^2} \hat{\mathbf{R}}^2,
\end{align}
must also be transformed. In 
Hund's case a representation, $\hat{\mathbf{R}} = \hat{\mathbf{J}} - \hat{\mathbf{L}} - \hat{\mathbf{S}}$ is the nuclear rotational angular momentum in the body-fixed frame. While $\hat{\mathbf{J}}$ is diagonal on $\mathbf{U}(r)$ and $\hat{\mathbf{L}}$ is irrelevant for our $\Sigma$ system, the $\Omega$ transformation of $\hat{\mathbf{S}}$ makes this property bond-length dependent due to spin-orbit mixing.
This is an important difference with the $\Lambda S$ basis, where the matrix elements of the spin operator  $\hat{\mathbf{S}}$ are constant by construction. 

\begin{figure}
    \centering
    \includegraphics[width=0.7\linewidth]{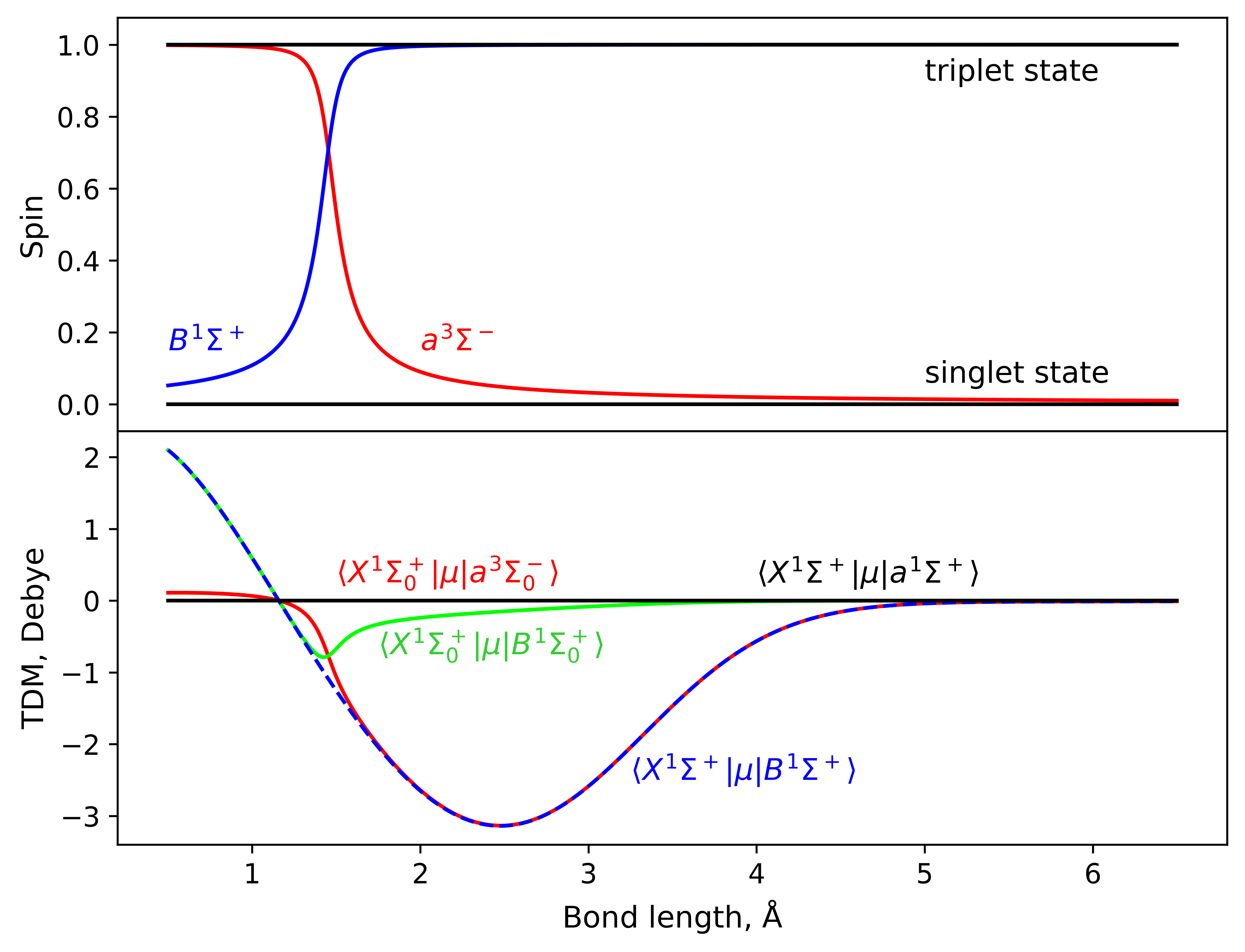}
    \caption{Illustrations of the spin eigenvalues (top) and TDMCs (bottom) as a function of bond length in the $\Omega$-representation. SOC-induced mixing swaps spin multiplicities between the $a$ and $B$ states, highlighting the emergence of a transition dipole moment. The corresponding constant spins in the $\Lambda S$ representation are also shown.}
    \label{fig:Omega_spin}
\end{figure}

Figure~\ref{fig:Omega_spin} illustrates the $r$-dependent behaviour of $\hat{\mathbf{S}}$  in the $\Omega$ representation: at short bond lengths, the $a\,^3\Sigma^-$ and $X\,^1\Sigma^+$ states retain their singlet and triplet character, respectively, but at larger $r$, strong SOC leads to mixing and spin character inversion across the avoided crossing. This spin evolution enables the emergence of the transition dipole moment due to this mixing, where both states acquire significant singlet-triplet character and satisfy spin selection rules.

The $B\,^1\Sigma^+ -X\,^1\Sigma^+$ transition dipole moment (TDM) is the only non-zero transition dipole in the $\Lambda S$ representation. Upon transformation to the $\Omega$ representation, an effective TDM between the $a\,^3\Sigma^-$ and $X\,^1\Sigma^+$ states is induced. Figure~\ref{fig:Omega_spin} illustrates the TDMCs in both representations, revealing the emergence of a non-zero TDM for the spin-forbidden $a \rightarrow X$ transition due to SOC-driven mixing.


As a quick reflection on the resulted transformation, we would like to note that although the $\Omega$ representation offers physical insight into SOC-driven mixing, it complicates the spectroscopic model (see also discussion in Ref.~\citenum{10TaKlKr.KCs}). Properties like spin and angular momentum become geometry-dependent, introducing non-trivial topologies in dipole moments and other observables. Small changes in the potential topology can lead to large variations in these properties. By contrast, the $\Lambda S$ representation is simpler and more practical: spin remains fixed, SOC is explicitly included, and all spin properties can be treated analytically, making it preferable for high-accuracy modelling.

\section{Computation of the Rovibronic Spectrum}

With the spectroscopic model defined above, our effectively complete rovibronic basis, either in the $\Lambda S$ or $\Omega$ representation, is used to construct the fully coupled Hamiltonian of Eq.(\ref{eq:diatomic_schrodinger_equation_duo}). The Hamiltonian is then diagonalised to yield a set of rovibronic energies and wavefunctions. A full non-adiabatic module has been implemented in our rovibronic code \duo\ \citep{Duo} to incorporate the NACs arising from the transformation of the vibrational nuclear kinetic energy. All functionality of \duo, previously implemented exclusively in the $\Lambda S$ representation, has now been extended to operate in the $\Omega$ representation as well, where all terms of the Hamiltonian are transformed.

Convergence in the sinc-DVR vibrational basis was achieved using a large contracted set of 1750 vibrational wavefunctions over a grid of 2001 points. This same basis was used for the $\Lambda S$ calculations. Rovibronic spectra were computed up to $J=70$ for all electronic states, with full inclusion of non-adiabatic couplings and bond-length-dependent spin-angular momentum terms in the $\Omega$-representation solution.

\begin{figure}
    \centering
    \includegraphics[width=0.7\linewidth]{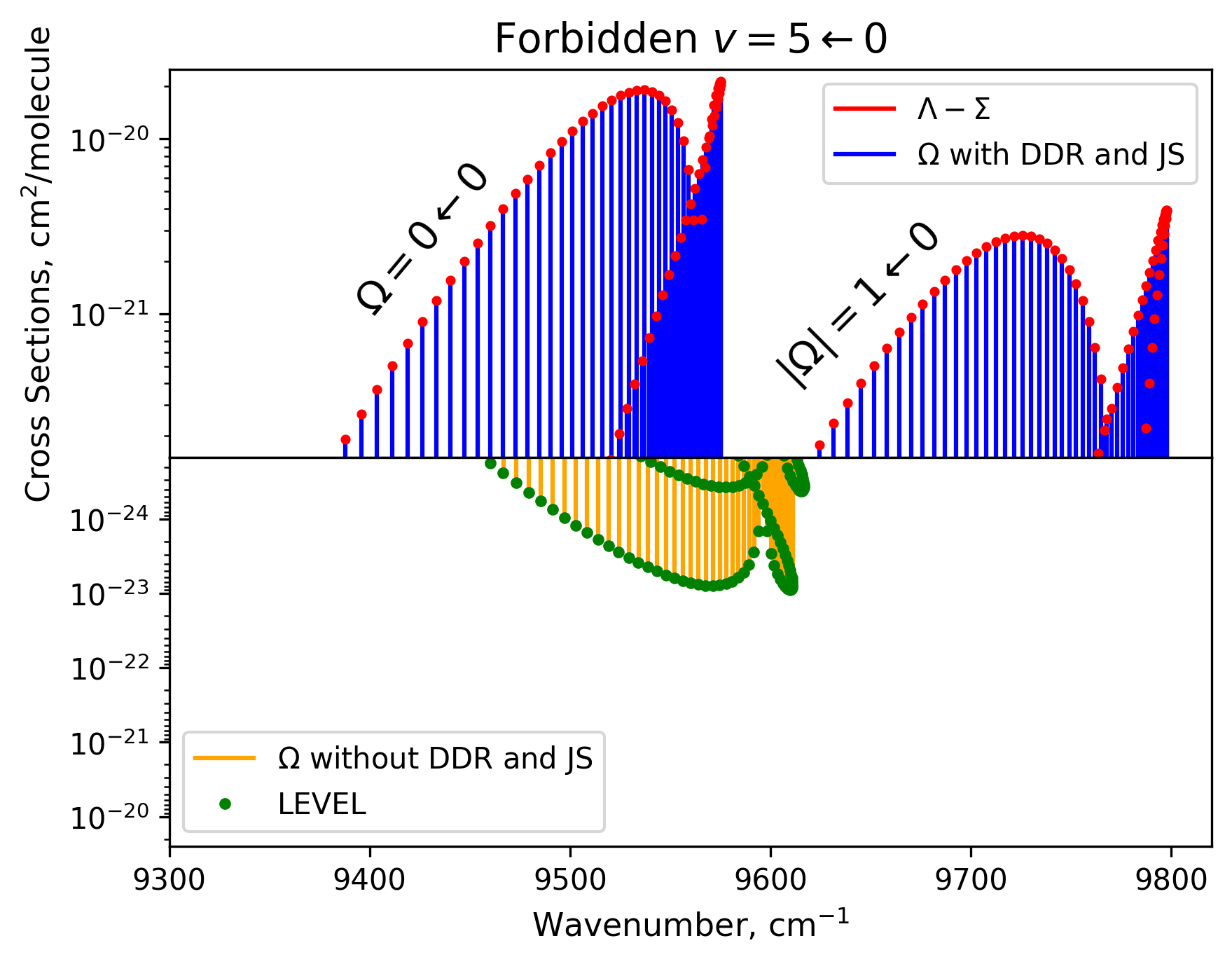}
    \caption{Comparison of our computed $a\,^3\Sigma^-\leftarrow X\,^1\Sigma^+$ $v=5\leftarrow0$ forbidden band intensities. The upper panel compares $\Lambda S$ (red) and $\Omega$ (blue) representations, showing their equivalence when all coupling terms are included. The lower panel compares our approximate \Duo\ $\Omega$-calculation (orange) with a single-state approximation computed with \textsc{Level} (green dots).}
    \label{fig:LEVEL_spectrum_comparison}
\end{figure}

Cross-sections for the resulting rovibronic stick-spectrum are then computed at a temperature of 298~K in both the $\Lambda S$ and $\Omega$ representations. Figure \ref{fig:LEVEL_spectrum_comparison} (upper panel) compares the forbidden $a\,^3\Sigma^-\leftarrow X\,^1\Sigma^+$ $v=5\leftarrow 0$ band intensities, which are seen to agree in both $\Lambda S$ (red sticks) and $\Omega$ (blue sticks) calculations, where the slight difference in the band intensity we attribute to convergence issues. 

To test the widely used single-state approximation, which omits non-adiabatic effects, we use the well-established \textsc{Level} program as a reference, which requires this approximation. We focus on the accuracy of line intensities only, since these cannot be empirically refined like rovibronic energies. Thus, we compute the $a\,^3\Sigma^-\leftarrow X\,^1\Sigma^+$ spin-forbidden band intensities with \textsc{Level} using the $\Omega$-representation PECs and TDMs from Figs~\ref{fig:Omega_LS_model_comp} and \ref{fig:Omega_spin}. These are illustrated for the $v=5\gets 0$ band and $T=296$~k in Fig.~\ref{fig:LEVEL_spectrum_comparison} (lower panel, orange sticks). 

As expected, reproducing the LEVEL results with \duo\ requires omitting all non-adiabatic terms, including NACs ($\mathbf{W}=\mathbf{W}^2=0$) and the bond-length dependence of the spin quantum number, $S(r)$. Fig.~\ref{fig:LEVEL_spectrum_comparison} illustrates the forbidden  $a\,^3\Sigma^-\leftarrow X\,^1\Sigma^+$ $v=5\leftarrow 0$ band spectrum (lower panel, green points) showing excellent agreement with the associated \textsc{Level} output. In contrast, comparison to the full $\Omega$ calculation (Fig.~\ref{fig:LEVEL_spectrum_comparison}, upper panel) reveals that omitting NACs and $S(r)$ leads to an underestimation of band intensity by 3 orders of magnitude with significant shifts in line positions. This demonstrates that the $\Omega$ representation with the single-state approximation, such as commonly used with \textsc{Level} is fundamentally inadequate for systems exhibiting strong mixing near the Franck-Condon region of the ground state. The computed band intensities can yield non-physical results if the non-adiabatic terms are not carefully treated.

\begin{figure}
    \centering
    \includegraphics[width=0.7\linewidth]{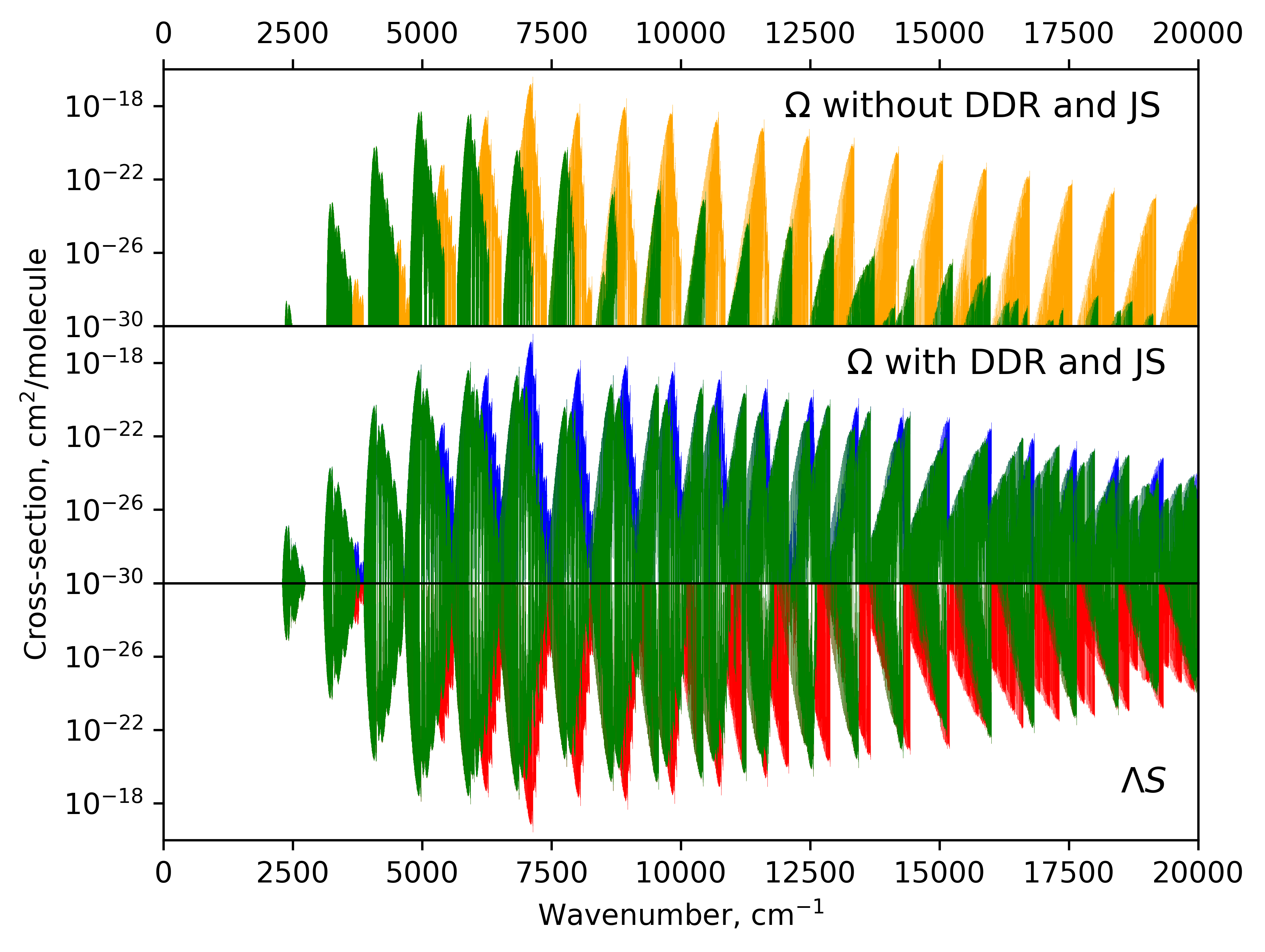}
    \caption{Computed band intensities for the studied system using \textsc{Duo}. The green lines show the forbidden $a\,^3\Sigma^-\leftarrow X\,^1\Sigma^+$ band, and the other colours (blue, red, and orange) show the full spectrum for the system of study.}
    \label{fig:full_band_comparison}
\end{figure}

Figure \ref{fig:full_band_comparison} presents a comprehensive comparison of the computed cross sections for the studied system over the full spectroscopic region. The bottom and middle panels show the total opacity for the fully coupled $\Lambda S-$ and $\Omega-$ representations, respectively. The total opacity is highlighted in different colours, with the forbidden $a\,^3\Sigma^-\leftarrow$$X\,^1\Sigma^+$ band highlighted in green. Our new \textsc{Duo} implementation confirms the exact equivalence of the total opacity calculated using both the $\Lambda S-$ and $\Omega-$ representations with all non-adiabatic terms treated, as expected. The top panel shows the same band intensities in the approximate single-channel $\Omega$ calculations when the DDR and spin-uncoupling terms are not treated. It is obvious that the approximate case does not accurately reproduce the global forbidden band intensities, and is thence unsuitable for high resolution applications -- at least in cases similar to the system we model here.

Figures~\ref{fig:LEVEL_spectrum_comparison} and \ref{fig:full_band_comparison} illustrate the main conclusion of this work: (1) equivalency between two representations based on the same underlying model is possible only when all non-adiabatic elements are included; (2) we reproduce the independent single channel \textsc{Level} results with \textsc{Duo} using a reduced, non-adiabatic-free model and (3) the results from the full and reduced treatments are drastically different both quantitatively (two bands vs one) and qualitatively (the intensities of the single state mode are over 3 orders of magnitude weaker). 
We conclude that the $\Omega$-representation, while often perceived as simpler, can be more complex and challenging to implement accurately. 
It is highly sensitive to the topology of property curves and often less practical than the spin-orbit coupled $\Lambda S$ representation. An exact decoupling scheme is not rigorously achievable, where \emph{simplification of one part of the Hamiltonian leads to the complication of another}.

In Table~\ref{tab:Omega_LS_energies_lifetimes}, we provide a more qualitative comparison of the computed energy terms and radiative lifetimes  between the fully coupled $\Omega$ and $\Lambda S$ representations as well as with the single-channel $\Omega$ representation reinforcing that 
the single-state channel  approximation in the $\Omega$-representation is not sufficient to yield accurate lifetimes, line positions, and line intensities for the systems of the presented type.

\begin{table*}
\footnotesize
    \centering
    \caption{Energy terms ($E$ [cm$^{-1}$]) and radiative lifetimes ($\tau$ [s]) for the lowest 10 vibrational $J=0$ $+$ states of the studied system. The 'I' subscript refers to quantities computed in the $\Omega$-representation with the spin-uncoupling and DDR terms omitted from the \duo\ calculations. Exact equivalence between the fully coupled $\Omega$- and $\Lambda-S$ representations can be seen, where differences to the approximate energies and lifetimes are given in the final two columns.}
    \label{tab:Omega_LS_energies_lifetimes}
    \begin{adjustbox}{angle=90}
    \begin{tabular}{rrrrcrcrrcrrr}
        \hline\hline 
        \multicolumn{5}{c}{$\Omega$ representation} && \multicolumn{4}{c}{$\Lambda-S$ representation} && \multicolumn{2}{c}{} \\
        \cline{1-6} \cline{8-11} 
$E$  & $\tau$ & $E$(I) & $\tau$(I) & State   &  $v$  &  &  $E$  &  $\tau$   &  State   &    $v$  & $\Delta E$ & $\tau_I/\tau$ \\
        \cline{1-6} \cline{8-11}
 4962.888218 & 9.30E-03  &  4962.142559  & 8.33E-03 & $a\,^3\Sigma^-_0$ & 0  &  &  4962.888218 & 9.30E-03  & $a\,^3\Sigma^-$ & 0 &  0.75  & 1.12E+00 \\ 
 5920.231428 & 2.51E-03  &  5920.190070  & 2.57E-03 & $a\,^3\Sigma^-_0$ & 1  &  &  5920.231428 & 2.51E-03  & $a\,^3\Sigma^-$ & 1 &  0.04 & 9.78E-01  \\ 
 6860.795910 & 7.72E-04  &  6863.989096  & 1.12E-03 & $a\,^3\Sigma^-_0$ & 2  &  &  6860.795910 & 7.72E-04  & $a\,^3\Sigma^-$ & 2 & -3.19  & 6.89E-01 \\ 
 7782.564891 & 3.09E-04  &  7792.383612  & 5.51E-04 & $a\,^3\Sigma^-_0$ & 3  &  &  7782.564891 & 3.09E-04  & $a\,^3\Sigma^-$ & 3 & -9.82  & 5.60E-01 \\ 
 8683.786984 & 1.53E-04  &  8703.796425  & 2.97E-04 & $a\,^3\Sigma^-_0$ & 4  &  &  8683.786984 & 1.53E-04  & $a\,^3\Sigma^-$ & 4 & -20.01  & 5.15E-01 \\ 
 9563.622535 & 9.75E-05  &  9596.312801  & 2.21E-04 & $a\,^3\Sigma^-_0$ & 5  &  &  9563.622535 & 9.75E-05  & $a\,^3\Sigma^-$ & 5 & -32.69  & 4.41E-01 \\ 
10422.297031 & 1.19E-04  & 10468.048772  & 3.16E-04 & $a\,^3\Sigma^-_0$ & 6  &  & 10422.297031 & 1.19E-04  & $a\,^3\Sigma^-$ & 6 & -45.75  & 3.77E-01 \\ 
11260.671233 & 8.58E-04  & 11317.762688  & 2.25E-03 & $a\,^3\Sigma^-_0$ & 7  &  & 11260.671233 & 8.58E-04  & $a\,^3\Sigma^-$ & 7 & -57.09  & 3.82E-01 \\ 
12079.741723 & 6.56E-04  & 12145.297828  & 1.90E-02 & $a\,^3\Sigma^-_0$ & 8  &  & 12079.741723 & 6.56E-04  & $B\,^1\Sigma^+$ & 6 & -65.56  & 3.45E-02  \\ 
12880.408956 & 4.68E-04  & 12951.396680  & 2.86E-02 & $a\,^3\Sigma^-_0$ & 9  &  & 12880.408956 & 4.68E-04  & $B\,^1\Sigma^+$ & 7 & -70.99  & 1.63E-02  \\ 
13663.525626 & 3.29E-04  & 13737.089066  & 7.02E-01 & $a\,^3\Sigma^-_0$ & 10 &  & 13663.525626 & 3.29E-04  & $B\,^1\Sigma^+$ & 8 & -73.56  & 4.68E-04 \\ 
\hline
    \end{tabular}
    \end{adjustbox}
\end{table*}

\section{Discussion}

Transforming rovibronic Hamiltonians  from the $\Lambda S$ (Hund’s case a) basis to the adiabatic $\Omega$ representation is often used to eliminate SOC and use  single-state treatments.  

We have demonstrated  using a diatomic benchmark system, where the theory can be rigorously tested and the results are directly applicable to high-accuracy sciences (e.g. ultracold physics), that this perceived simplification is not free: the transformation necessarily introduces non-adiabatic effects from the nuclear kinetic energy operator. Neglecting non-adiabatic effects resulted in significant errors in both predicted intensities and energies. The errors of the single-state approximated $\Omega$-representation are likely to persist in polyatomic systems, in which analogous electron-nuclear couplings are of comparable magnitude and known to be of spectroscopic importance.

While accounting for non-adiabatic effects in polyatomics is often computationally intractable due to the complexity of the kinetic energy operators, single-state approximations remain the only viable route. Particularly in high-resolution spectroscopy, this approximation is only reliable when the interacting electronic states are sufficiently isolated (as argued by Hellman-Feynman Theorem) or when avoided crossings occur well outside the spectroscopic region of interest (e.g. the Franck-Condon of the ground state). Besides, errors in rovibronic energies can sometimes be masked by empirical refinement of of the model through the ability to `absorb' errors into the potential energy surfaces. However, intensities and radiative lifetimes remain sensitive because dipole moments are typically treated \ai. 
This being said, for computing spin-forbidden band intensities only a portion of the signal is reproduced in the single-state $\Omega-$representation. Transformation of the spin-uncoupling term allows $\Omega$ changing transitions which are otherwise not reproduced from the `uncoupled' potential energy surfaces alone. Thus, even in the best approximate cases one needs to introduce further terms to construct a full spectroscopic signal.

This work aims to debunk the common belief that the uncoupled $\Omega$-representation can accurately model forbidden bands. It urges the community to stop using single-state models blindly without justifications and define the limits within which such approximations are valid or to consider including non-adiabatic effects. Our practical recommendation is to always inspect adiabatic effects and transition dipoles in the $\Omega$ frame for sharp features near the Franck-Condon window;  if red flags appear, switch to a minimally coupled treatment ($\Omega$ with induced NACs, or $\Lambda S$ with explicit SOC). A possible mitigation of the non-adiabatic effects can be achieved using perturbation theory. We also recommend to always provide a short diagnostic summary (e.g., nearest avoided crossing, SOC scale, expected intensity borrowing). This helps others judge transferability and reproducibility.


Our analysis focuses on two-state scenarios and diatomic benchmarks, enabling rigorous proofs and clean error attribution. Systems with multiple nearby states, strong rotational couplings, or additional vibronic interactions will require extended diagnostics and potentially higher-rank NAC treatments. For polyatomics, scalable representations of the kinetic-energy couplings remain an algorithmic bottleneck; reduced-space variational strategies and perturbative hybrids are promising pathways. Incorporating these ingredients into widely used workflows should be a community priority, especially for high-accuracy intensity predictions.


To conclude, eliminating SOC via transformation to the $\Omega$ representation does not, by itself, simplify the rovibronic problem; it relocates the complexity into induced non-adiabatic terms. Accurate prediction of energies, intensities, and lifetimes, crucial for high-accuracy applications, especially for spin-forbidden bands, requires these terms to be included. Our derivations establish the exact equivalence conditions between $\Omega$ and $\Lambda S$ formulations, our computations quantify the resulting errors when terms are omitted, and our diagnostics and remedies provide a practical path to reliable predictions within existing spectroscopy pipelines. For general computational and theoretical chemistry, the broader lesson is clear: unitary ``simplifications'' must be accompanied by an accounting of the physics they move elsewhere in the Hamiltonian.


\section*{Acknowledgements}

This work was supported by the European Research Council (ERC) under the European Union’s Horizon 2020 research and innovation programme through Advance Grant number 883830, the STFC grants ST/Y001508/1 and UKRI/ST/B001183/1.
We thank Jonathan Tennyson for supporting this work and providing valuable suggestions.

\bibliography{bib/journals_phys,bib/diabatisation,bib/abinitio,./bib/programs,./bib/methods,bib/Books,bib/PSe,bib/PO,bib/NSe,bib/YbBr,bib/SO,bib/SO+,bib/AgH,bib/BaLi+,bib/LuF,bib/MgCl+,bib/KCs,bib/SiO+,bib/CaCs,bib/BeSe,bib/CH-,bib/AlI,bib/KRb,bib/SrYb,bib/Rb2,bib/omega_rep,bib/diatomic,bib/thesis,bib/SO2} 

\end{document}